\documentstyle[aps,prb,epsfig]{revtex}
\begin{document}
\draft
\author{M. Kiselev$^1$,  K.Kikoin$^2$ and R. Oppermann$^1$}
\address{$^1$Institut f\"ur Theoretische Physik, Universit\"at W\"urzburg, D-97074, Germany,\\
$^2$Ben-Gurion University of the Negev, Beer-Sheva 84105, Israel}
\date{\today}
\title{ 
Ginzburg-Landau functional for nearly antiferromagnetic\\  perfect and disordered Kondo lattices}
\maketitle
\begin{abstract}
Interplay between Kondo effect and trends to antiferromagnetic and spin glass ordering 
in perfect and disordered bipartite Kondo lattices is considered. Ginzburg-Landau equation 
is derived from
the microscopic effective action written in three mode representation
(Kondo screening, antiferromagnetic correlations and spin liquid correlations).
The problem of local constraint is resolved by means of Popov-Fedotov representation 
for localized spin operators. It is shown that the Kondo screening enhances the trend to a
spin liquid crossover and suppresses antiferromagnetic ordering in perfect Kondo 
lattices and spin glass ordering in doped Kondo lattices. The modified Doniach's
diagram is constructed, and possibilities of going beyond the mean field approximation
are discussed.\\
\mbox{}
\end{abstract}
\section{Introduction}

The Kondo lattice (KL) systems are famous with their unusual electronic and magnetic properties,
including giant effective masses observed in thermodynamic and de Haas-van Alphen measurements 
\cite{spring}, unconventional superconductivity \cite{sup} and fascinating variety of magnetic properties
\cite{mag}. The overwhelming majority of metallic KL systems demonstrate antiferromagnetic (AFM)
correlations, and one can meet all types of AFM order:  localized spins in ${\rm U_2Zn_{17},
UCd_{11}, CeIn_3}$ \cite{mag};
quadrupole ordering in ${\rm CeB_6}$ \cite{quad};
interplay between localized and itinerant excitations in several U- and Ce-based compounds \cite{iti};
puzzling magnetic order of tiny moments in ${\rm UPt_3,URu_2Si_2, UNi_2Al_3}$ that, 
however, rapidly transform in usual large localized moments with doping \cite{fran}; 
quantum phase transition in ${\rm CeCu_{6-x}Au_x}$ \cite{lohn}; 
fluctuation-type dynamical ordering in ${\rm U(Pt_{1-x}Pd_x)_3}$ \cite{viss}; 
short-range magnetic correlations without long range order but with astonishingly wide 
temperature interval of critical behavior in ${\rm CeCu_6}$ and ${\rm CeRu_2Si_2}$ \cite{nord}. 
The list is by no means exhaustive.
The superconducting
state in most cases also coexists with antiferromagnetism, and, apparently, Cooper
pairing in KL is mediated by magnetic fluctuations \cite{sup,mathur}. 
The dominant contribution to the low-temperature thermodynamic
is also given by spin degrees of freedom \cite{var85,kkp}. 

On the other hand, all low-temperature characteristics of KL are scaled 
by  a Kondo temperature $T_K$.
These characteristics are Fermi-like, but the energy scale of "fermion" 
spectrum is compressed with a factor 
$T_K/\varepsilon_F$ relative to a conventional electron Fermi liquid  \cite{mag}.                          
Apparently, the AFM correlations due to Ruderman-Kittel-Kasuya-Yosida
(RKKY) interaction $I$ partially suppressed by intra-site Kondo effect  should be treated  
as a background for all  
unusual properties of Kondo lattices,  so the main theoretical challenge  is to find a scenario
of crossover from a high temperature regime of weak interaction (scattering) between localized spins
and conduction electron Fermi liquid to a low-temperature strong coupling regime where the spins 
lose their localized nature and are confined into unconventional quantum liquid involving spin degrees 
of freedom of conduction electrons. 

On the phase diagram of disordered KL more exotic possibilities arise, such as non Fermi liquid 
regime observed, e.g. near the $T=0$ quantum critical point in ${\rm Y_{1-x}U_xPd_3}$ (see, e.g., 
Ref. \onlinecite{yupd3_1}). In this family of ternary alloys the spin glass (SG) behavior was
discovered in a U concentration range $0.3<x<0.5$ with a freezing temperature $T_f$ 
growing monotonically with $x$ \cite{yupd3_3}. Among other ${\rm U}$- based 
heavy fermion compounds with SG behavior
${\rm URh_2Ge_2}$ \cite{urhge_1}, ${\rm U_2Rh_3Si_5}$ \cite{urhsi_1}, 
${\rm U_2PdSi_3}$
\cite{updsi_1} should be mentioned.
The effects of "Kondo disorder" were reported for ${\rm UCu_{5-x}Pd_x}$ in Ref. \onlinecite{ucupd_1}. 
Later on the competition between RKKY  and Kondo exchange for disordered ${\rm Ce}$ alloys
was discovered  experimentally (see Refs. \onlinecite{cenicu_1}-\onlinecite{cecogesi_1}).
The magnetic
phase diagram of ${\rm CeNi_{1-x}Cu_x}$ exhibits change of magnetic ordering
from AFM to FM at $x=0.8$, whereas for $0.2<x<0.8$ the SG state appears 
only at high temperatures above the FM order. Apparently, the Kondo interaction could be
considered as the mechanism leading to reduction of magnetic moments because 
increasing ${\rm Ni}$ contents effectively reduces the strength of indirect exchange interaction,
and then, a larger temperature 
stability range of the SG phase appears (see Refs.  \onlinecite{cenicu_1}-\onlinecite{cenicu_2}). 

The competition between the one-site Kondo type 
correlations and the indirect inter-site exchange is visualized in a Doniach's diagram where all
possible phase transition and crossover energies are plotted as a function of a "bare" coupling parameter 
$\alpha=J/\varepsilon_F$ that  characterizes  exchange interaction between the spin and electron
subsystem in KL \cite{doni}. Only Kondo screening and RKKY coupling were competing in 
original Doniach's diagram. Later on it was noticed that   
the trend to spin liquid (SL) ordering  is the third
type of correlation which modifies essentially the magnetic phase diagram of
KL in a critical region $T_K\sim I$ of the Doniach's diagram 
\cite{rvb,kkm94,coq}. 

In this paper we present a high temperature mean-field description of transitions from a
paramagnetic state to correlated spin states in KL, which does
not violate the local constraint for spin-fermion operators. 
We use the Popov-Fedotov representation
of spin operators \cite{pop} to construct the effective action for 
KL. In this representation the local constraint is satisfied
automatically. We consider the mutual influence of various  order parameters
(Kondo, AFM, SL and SG correlation functions) 
and derive a Ginzburg Landau functional (Section II).
On the basis of this functional we construct 
generalized Doniach's diagrams that take into account all interplays. The Doniach's diagram
for a  perfect KL is presented in Section III and the influence of Kondo screening on the 
SG diagram in disordered KL is considered in Section IV.  

All existing theories appeal to mean-field approximations that violate
the local gauge invariance both in Kondo and SL channel \cite{lark}.
As a result, fictitious second order phase transitions  from a free spin (paramagnetic)
state to a confined spin (Kondo singlet or resonating valence bond 
SL) state arise in spite of the fact that neither symmetry is violated by these 
transformations. New approach allows us to get rid of assumption about Kondo-type "condensate"
within a framework of mean field approach. To eliminate fictitious phase transition
to a SL state  one should refuse from a mean field approach to the SL mode. 
We offer scenarios of continuous crossover from a paramagnetic state of localized spins to
SL state, where the interplay between critical AFM fluctuations and Kondo screening
clouds in KL result in "fermionization" of spin excitations at low temperatures (Section V).
In section VI the interrelations between the theory and real heavy fermion systems is briefly 
discussed.  

\section{Derivation of effective action}

The  Hamiltonian of the KL model is given by 
\begin{equation}
H=\sum_{k\sigma}\varepsilon_k c^\dagger_{k\sigma}c_{k\sigma}+
J\sum_{j}\left({\bf S}_j{\bf s}_j+\frac{1}{4}N_jn_j
\right)
\label{2.1}
\end{equation}
Here the local electron and spin density operators for conduction electrons 
in a site $j$ are defined as 
\begin{equation}
n_j=\sum_{j\sigma}c^\dagger_{j\sigma}c_{j\sigma},~~~ 
{\bf s}_j=\sum_{\sigma}\frac{1}{2}c^\dagger_{j\sigma}
{\hat\tau}_{\sigma\sigma'}c_{j\sigma'}, 
\label{2.2}
\end{equation}
where ${\hat\tau}$ are the Pauli 
matrices and $c_{j\sigma}=\sum_k c_{k\sigma}\exp (ikj)$.
The SG freezing is possible if additional quenched randomness
of inter-site exchange $I_{jl}$ between the localized spins arises. 
This disorder is described by 
\begin{equation}
H'=\sum_{jl}I_{jl}({\bf S}_j{\bf S}_l).
\label{2.3}
\end{equation}

We start with a perfect Kondo lattice. The spin correlations in KL are
characterized by two energy scales, i.e., $I\sim \ J^2/\varepsilon_F,$ and 
$\Delta_K\sim \varepsilon_F\exp(-\varepsilon_F/J)$ 
(the inter-site indirect exchange of the Ruderman-Kittel-Kasuya-Yosida (RKKY) type
and the Kondo binding energy, respectively). At high enough temperatures the
localized spins are weakly coupled with the electron Fermi sea 
having the Fermi energy $\varepsilon_F$, so that the magnetic
response of a rare-earth sublattice of KL is of paramagnetic Curie-Weiss type.
With decreasing temperature either crossover to a strong-coupling 
Kondo singlet regime occurs at $T \sim \Delta_K$ or the phase transition 
to an AFM state occurs at $T=T_N \sim zI$ where $z$ is a coordination number
in KL. If $T_N \approx \Delta_K$ the interference between two trends result
in decreasing of both characteristic temperatures 
or in suppressing one of them. As was noticed in Refs
\onlinecite{kkm94,kkm97}, in this case the SL correlations with 
characteristic energies $\Delta_s\sim I$ may overcome the AFM correlations,
and the spin subsystem of KL can condense in a SL state yet in a region 
of weak Kondo coupling.

To describe all three modes in a unified way one should derive a  free energy
functional ${\cal F}(T)$ in a  region of 
$T > (T_N, \Delta_K, \Delta_s)$.
First, we integrate out the highest energies $\sim \varepsilon_F.$
Here and below we use the dimensionless coupling constants 
$J\to J/\varepsilon_F$, $I\to I/\varepsilon_F$, etc. 
Since we are still in a weak coupling limit of Kondo-type scattering,
we may  confine 
ourselves by the standard high temperature renormalization of the one site
coupling $J\to \tilde{J}(T)= 1/\ln(T/\Delta_K)$ and the second-order equation of 
perturbation theory in $J$ for RKKY interaction. As a result, one comes to an 
effective Hamiltonian
\begin{equation}
\widetilde{H} =\sum_{k\sigma}\varepsilon_k c^\dagger_{k\sigma}c_{k\sigma}
+\widetilde J\sum_j{\bf s}_j{\bf S}_i -I\sum_{jl}{\bf S}_j{\bf S}_l +g
h\sum_i(-1)^j S^z_j
\label{2.6}
\end{equation}  

Here all energies are measured in $\varepsilon_F=1$ units, and an infinitesimal staggered magnetic 
field is introduced that respects 
the symmetry of magnetic bipartite lattice in AFM case ($\varepsilon_F$ is restored in further
calculations wherever it is necessary). 

To calculate the spin part of free energy ${\cal F}_s(T)=-T\ln {\cal Z}_s$
we represent the partition function ${\cal Z}$ in terms of path integral.
The spin subsystem is described by means of Popov-Fedotov trick \cite{pop}
\begin{equation}
{\cal Z}_s
={\tt Tr}\;e^{-\beta H}=i^N {\tt Tr}\;e^{-\beta(H+i\pi N^f/(2\beta))},\
\label{2.7}
\end{equation}
Here $\beta=T^{-1},$  $N$ is the number of unit cells, 
$N^f=\sum_j N_j^f,$
and the
spin $S=1/2$ operators are represented as bilinear combinations of
fermion operators 
\begin{equation}
S_j^z=(f^\dagger_{j\uparrow}f_{j\uparrow}- 
f^\dagger_{j\downarrow}f_{j\downarrow})/2,~
S_j^+=f^\dagger_{j\uparrow}f_{j\downarrow},~
S_j^-=f^\dagger_{j\downarrow}f_{j\uparrow}.
\label{2.4}
\end{equation}
These operators obey the constraint
\begin{equation} 
N^f_j=\sum_\sigma f^\dagger_{j\sigma}f_{j\sigma}=1. 
\label{2.5}
\end{equation}
In accordance with Ref. \onlinecite{pop},
the Lagrange term with 
a fixed imaginary chemical potential $-i\pi T/2$ is added to the Hamiltonian 
(\ref{2.1}).
We use the path integral representation for the partition function,
\begin{equation}
\frac{\cal Z}{\cal Z}_0=\frac{\int D\bar c D c D\bar f D f \exp{\cal A}}{
\int D\bar c D c D\bar f D f\exp{\cal A}_0}
\label{2.7a}
\end{equation}
Then the Euclidean action for the KL is given by
$${\cal A}={\cal A}_0 - \int_0^\beta d\tau{\cal H}_{int}(\tau)$$
\begin{equation}
{\cal A}_0={\cal A}_0[c,f] =\int_0^\beta d\tau \sum_{j\sigma}\left\{
\bar{c}_{j\sigma}(\tau)[\partial_\tau -\varepsilon(-i\nabla)+\mu]
c_{j\sigma}(\tau)+
\bar{f}_{j\sigma}(\tau)(\partial_\tau - i\pi T/2) f_{j\sigma}(\tau)\right\}
\label{2.8}
\end{equation}

Following the Popov-Fedotov procedure, the imaginary chemical potential is 
included in discrete Matsubara frequencies for semi-fermion operators 
$f_{j\sigma}$. As a result the Matsubara frequencies are determined as
$\omega_m=2\pi T(m+1/4)$ for spin semi-fermions and 
$\epsilon_n=2\pi T(n+1/2)$ for conduction electrons. 
In terms of temperature Green's function the Euclidian action has the form
\begin{eqnarray}
{\cal A} = {\cal A}_0 +{\cal A}_{int} & = & \sum_{k \sigma} \bar{c}_{k\sigma} 
G_0^{-1}(k)c_{k\sigma}
+\sum_{j \sigma} \bar{f}_{j\sigma}(\omega_n)
D_{0\sigma}^{-1}(\omega_n)f_{j\sigma}(\omega_n)\nonumber \\
& + & \frac{\widetilde{J}}{2} \sum_{j\sigma\sigma'}\sum_{\varepsilon_m,\omega_n}
\bar{c}_{j\sigma}(\varepsilon_1)f_{j\gamma'}(\omega_2)
\bar{f}_{j,\sigma'}(\omega_1)c_{j,\sigma'}(\varepsilon_2)
\delta_{\varepsilon_1-\varepsilon_2,\omega_1-\omega_2}
\nonumber \\
& + &
I \sum_{jl,\sigma\gamma}\sum_{\omega_n}\bar{f}_{j\sigma}(\omega_1)
\hat{\tau}_{\sigma\sigma'} f_{j,\sigma'}(\omega_2)
\bar{f}_{l,\gamma}(\omega_3)\hat{\tau}_{\gamma\gamma'}
f_{l\gamma'}(\omega_4)\delta_{\omega_1-\omega_2,\omega_3-\omega_4}.
\label{2.9}
\end{eqnarray}
Here the Green's functions (GF) for bare quasiparticles are  
\begin{equation}
G_0(k,i\epsilon_n)=\frac{1}{i\epsilon_n -\varepsilon_k
+\mu},\;\
D^\nu_{0\sigma}(i\omega_m)=
\frac{1}{i\omega_m -\sigma g h^\nu/2}
\label{2.10}
\end{equation}
($\nu$ is the index of magnetic sublattice that defines the direction of 
staggered magnetic field).

The first interaction term in this equation is responsible for {\it low-energy}
Kondo correlations, and we will treat it in conventional manner \cite{ren}.
In the RKKY term two modes should be considered, i.e. the local mode of
AFM fluctuations \cite{bok,kop1} and the 
nonlocal spin liquid correlations \cite{kop1,kkm97a}. 
We adopt for these modes the 
Neel type antiferromagnetism and resonating valence bond 
(RVB) type spin liquid state, respectively. 
In accordance with the general path integral approach to KL, 
we first integrate over
fast (electron) degrees of freedom. Then in the sf-exchange contribution to the
action (\ref{2.9}) we are left with the auxiliary field $\phi$ with statistics 
complementary to that of semi-fermions
\cite{ft1}. 
 The spin correlations in the inter-site
RKKY term are treated in terms of vector Bose fields ${\bf Y}$ (AFM mode) and 
scalar field $W$ (spin liquid RVB mode). 
As a result, ${\cal A}_{int}$ is represented 
by a following expression
\begin{equation}
{\cal A}_{int}   =  -\frac{2}{\widetilde J}{\rm Tr}|\phi|^2 - 
{\rm Tr}\frac{1}{I_{\bf q}}{\bf Y}_{\bf q}{\bf Y}_{\bf -q} -   
{\rm Tr}\frac{1}{I_{{\bf q}_1-{\bf q}_2}}W_{{\bf Pq}_1}W_{{\bf Pq}_2}
-  {\rm Tr}\bar{f}_{j\sigma}\phi_j  G_0({\bf r}) \bar{\phi}_lf_{l\sigma}~.
\label{2.11}
\end{equation}   
When making a Fourier transformation for non-local spin liquid 
correlations (the third term in the r.h.s.) we introduced the coordinates 
${\bf R}=({\bf R}_j+{\bf R}_l)/2$ and ${\bf r}={\bf R}_j-{\bf R}_l$ 
for RVB field, so that ${\bf P},{\bf q}$ are the corresponding momenta. 
Below we assume ${\bf P}=0$ and omit it in notations for the SL mode,
$W_{0{\bf q}}\equiv W_{\bf q}$.

Consequent mean-field approach demands introduction of three "condensates",
i.e. three time independent c-fields for 
Kondo coupling, AFM coupling and SL coupling, respectively, that arise
as a consequence of a saddle-point approximation for all three modes.
For example, the mean-field description of the interplay between the
Kondo and RVB couplings was presented in \cite{rvb,coq}. 
The undesirable consequence of this approximation is  violation of the
electromagnetic $U(1)$ gauge invariance, when the electrical charge
is ascribed to initially neutral spin fermion field $f$ (see, e.g. 
Ref.\onlinecite{kkp}).  
According to a
scenario offered in \cite{kkm94,kkm97a}, there is no necessity in 
introducing the mean-field saddle point for Kondo coupling because the
transition to a correlated spin state occurs at $T>T_K$. In this case 
the one site Kondo correlations suppress the Neel phase transition 
(reduce $T_{N}^0\to T_N$) in favor of spin liquid state with a characteristic
crossover temperature $T^*>T_N$. So we  refrain from using the 
saddle point approximation for the field $\phi$ but still use it for the
fields ${\bf Y}$ and $W$. 

To compactify the equation for the action ${\cal A}$ we introduce a spinor 
representation for semi-fermions
$$
\bar{F}_{\bf p}=\left(\bar{f}_{{\bf p}\uparrow}\;\bar{f}_{{\bf p}\downarrow}\;
\bar{f}_{{\bf p+Q}\uparrow}\;\bar{f}_{{\bf p+Q} \downarrow}
\right)~,
$$
and the 
following definition of the Fourier transform of inverse 
semi-fermionic Green's function 
\begin{eqnarray}
D^{-1}_m(W_{\bf p},{\bf Y}_{\bf Q})=
\left(
\begin{array}{cccc}
i\omega_m-W_{\bf p} & 0         & Y^z_{\bf Q}         & Y^+_{\bf Q}\\
0         & i\omega_m-W_{\bf p} & Y^-_{\bf Q}         & -Y^z_{\bf Q}\\
Y^z_{\bf Q}       &  Y^+_{\bf Q}        & i\omega_m-W_{\bf p+Q} & 0\\
Y^-_{\bf Q}       & -Y^z_{\bf Q}        & 0           & i\omega_m-W_{\bf p+Q}
\end{array}
\right)~.
\label{2.12}
\end{eqnarray}
The same function in a lattice representation is presented in Appendix I.
This operator arises as a result of Hubbard-Stratonovich transformation 
decoupling the magnetic modes ${\bf Y}$ and the spin-liquid mode $W$.
Then the effective action ${\cal A}_s$ acquires the following form 
\begin{equation}
{\cal A}_s= {\rm Tr} \bar{F}D^{-1}_m F + {\cal A}_{int}. 
\label{2.13}
\end{equation}
Now we  integrate over semi-fermionic fields and obtain the effective action 
for a KL model 
\begin{equation}
{\cal A}_s = 
{\rm Tr} \ln\left(D_m^{-1}({\bf Y},W)+ \phi_{j}G_0({\bf r})\bar{\phi_l} \right)
-\frac{2}{\widetilde J}{\rm Tr}|\phi|^2 - 
{\rm Tr}\frac{1}{I_{\bf q}}{\bf Y}_{\bf q}{\bf Y}_{\bf -q} -   
{\rm Tr}\frac{1}{I_{{\bf q}_1-{\bf q}_2}}W_{{\bf q}_1}W_{{\bf q}_2}~.
\label{2.14}
\end{equation}
Here the argument $|\phi|^2$ appeared in the Green's function $D_m$ as
a result of integration of the last term in eq. (\ref{2.11}) over the 
semi-fermionic fields. 

In a mean-field approximation for two independent modes (neglecting 
renormalization due to Kondo scattering) eq. (\ref{2.14}) results in a 
free energy with two local minima reflecting two possible instabilities of 
high-temperature paramagnetic state relative to Neel and SL
states. To describe these instabilities one should  pick out the classic
part of Neel field
\begin{equation}
{\bf Y} =(\beta N)^{1/2} \frac{I_{\bf q}}{2}{\cal N}\delta_{\bf q,Q}\delta_{\omega,0}{\bf e}_z +
\widetilde{\bf{Y}}_{\bf q}
\label{2.15}
\end{equation} 
and use the eikonal approximation for the SL field
\begin{equation}
W_{\bf R,r}=I\Delta({\bf r})\exp (i\theta)~.
\label{2.15a}
\end{equation}
Here ${\cal N}=\langle Y^z_{\bf Q}\rangle$ is the staggered magnetization,
$\widetilde{\bf{Y}}_{\bf q}$ are the fluctuations around the mean-field \
magnetization,
${\bf Q}$ is the AFM vector for a given bipartite lattice, ${\bf e}_z$ is
the unit vector along the magnetization axis, 
$\Delta({\bf r})$ is the modulus of RVB field, and 
$\theta =({\bf r}\cdot{\bf A}({\bf R}))$ is the phase of this field.

As is known, in Heisenberg lattices for dimension $d>1$,
$T_N$ is higher than the temperature $T_{sl}$ of crossover to the SL state, 
so that the ordered magnetic phase is the Neel
phase. Due to Kondo fluctuations that screen dynamically local magnetic  
correlations and slightly enhance the inter-site semi-fermionic correlations,
 the balance between two modes is shifted towards spin liquid phase in 
a critical region of Doniach's diagram, $T_K\sim I.$ To show this we include
in the free energy the corresponding corrections induced by the last term in
Eq. (\ref{2.11}). As was mentioned above we refrain form using the mean-field
approach to Kondo field, so that the interplay between the Kondo mode and
two other modes is taken into account by including the Neel mean filed 
corrections to the semi-fermionic Green's function. Then instead of (\ref{2.12})
one has the following equation for $D^{-1}$:
\begin{eqnarray}
D^{-1}_m({\cal N}, \Delta)=
\left(
\begin{array}{cccc}
i\omega_m-\Delta I_{\bf q}& 0         & {\cal N}I_{\bf Q}/2         & 0\\
0         & i\omega_m-\Delta I_{\bf q} & 0         & -{\cal N}I_{\bf Q}/2  \\
{\cal N}I_{\bf Q}/2         &  0        & i\omega_m-\Delta I_{\bf q} & 0\\
0       & -{\cal N}I_{\bf Q}/2          & 0           & i\omega_m-\Delta I_{\bf q}
\end{array}
\right)~.
\label{2.18}
\end{eqnarray}
The next steps, i.e. calculations of fluctuation corrections to the stationary
point mean field solutions, can be performed by introducing the auxiliary 
self energies 
\begin{eqnarray}
M(\widetilde{\bf Y},\theta) & = & D_m^{-1}({\bf Y},W)-D_m^{-1}({\cal N},\Delta)
\nonumber\\
K_{\phi}(\omega_{n_1},\omega_{n_2}) & = & -T\sum_{\Omega}\phi_j(\omega_{n_1}-\Omega) G_0({\bf r},\Omega)
\bar{\phi}_l(\omega_{n_2}-\Omega)
\label{2.18a}
\end{eqnarray}
Then the effective action is approximated by the polynomial expansion
\begin{equation}
{\rm Tr} \ln \left(D_m^{-1}({\bf Y},W) +K_{\phi} \right) =
{\rm Tr} \ln D_m^{-1}({\cal N},\Delta) + {\rm Tr}
\sum_{n=1}^\infty\frac{(-1)^{n+1}}{n}\left\{D_m({\cal N},\Delta)
[M(\widetilde{\bf Y},\theta) + K_\phi]
\right\}^n
\label{2.18b}
\end{equation}
(the Fourier transform of diagonal part of the  Green's function $K_{\phi}$ is calculated in
Appendix I). 

Neglecting all fluctuations, i.e. retaining only the first term in the r.h.s. 
of Eq. (\ref{2.18b}) together with quadratic terms for AFM and SL modes (\ref{2.14}), 
one obtains the following expression for the free energy per lattice cell:
\begin{equation}
\beta{\cal F}({\cal N}, \Delta)=\frac{\beta z |I| {\cal
N}^2}{4}- \ln\left[2\cosh(\beta z I {\cal N}/2)\right]
+\frac{\beta  z|I| 
\Delta^2}{2}- \sum_{\bf q}\ln\left[2\cosh(\beta  I_{\bf q}\Delta)\right]
\label{2.16}
\end{equation}
($I_{\bf Q}=-I$).  
The standard self-consistent mean-field equations for order parameters are
obtained by differentiating eq. (\ref{2.16}) from condition of minima of the 
free energy. These are
\begin{equation}
{\cal N}= \tanh\left(\frac{I_{\bf Q}{\cal N}}{2T}
\right)
\label{2.16a}
\end{equation}
for the Neel parameter and
\begin{equation}
\Delta= -\sum_{\bf q} \nu({\bf q}) \tanh\left(\frac{I_{\bf q}\Delta}{T}
\right)
\label{2.16b}
\end{equation}
for the real part of the RVB order parameter. Here $ \nu({\bf q})= I_{\bf q}/I_0$.
The latter equation was first derived in Ref. \onlinecite{ban}. 

Then making the high temperature expansion 
of Eq. (\ref{2.16}), one obtains a Ginzburg-Landau (GL)
equation in the approximation of
two independent modes:
\begin{equation}
\beta{\cal F}({\cal N}, \Delta) =\frac{\beta |I| z{\cal N}^2}{4}\tau_N + c_N {\cal N}^4 +
\frac{\beta  |I|z 
\Delta^2}{2}\tau_{sl} + c_{sl}\Delta^4
\label{2.17}
\end{equation}
where $\tau_N=1-T_N/T$ and $\tau_{sl}=1-T_{sl}/T$. The temperatures of two magnetic
instabilities are determined as the temperatures of sign inversion in the coefficients
in quadratic terms of GL expansion $T_N=z|I|/2$ and $T_{sl}=|I|$. The 
forth order GL coefficients  $c_N$ and $c_{sl}$ are positive and 
depend only on temperature. 
Up to this point the theory is formulated for arbitrary dimension $d$. In fact, the dimensionality
enters the RKKY coupling parameter (see below) and determines the number of nearest neighbors $z$.
We consider $zI$ as a universal parameter in further calculations.

\section{Doniach's diagram revisited}

To describe the contribution of Kondo scattering to magnetic part
of the Doniach's diagram one should
 integrate ${\cal A}$ over auxiliary field $\phi$ and thus find the
Kondo corrections both to Neel and RVB instability points.
One should consider two cases: (i) $T_N>T_{sl}$ (Kondo corrections screen 
AFM magnetic moments), and (ii) $T_{sl}>T_N$ (Kondo corrections enhance
nonlocal RVB correlations).

(i) {\it Kondo screening of AFM order}. In this case one takes $\Delta=0$ in 
the Green's function (\ref{2.18}). Then adding the last term of eq. (\ref{2.11})
to the effective action and integrating over the semi-fermionic fields one gets
correction to the effective action in a form of polarization operators given
by the first diagram in Fig. 1a. 
\begin{figure}
\begin{center}
  \epsfxsize46mm \epsfbox{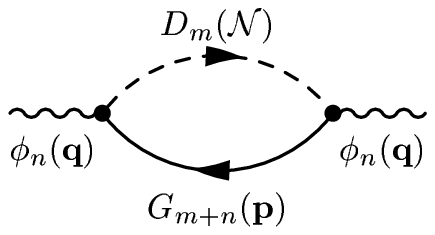}\hspace*{2cm}
  \epsfxsize46mm \epsfbox{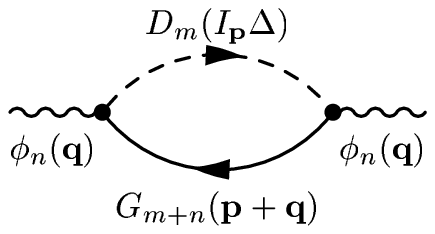}
a)\hspace*{7cm} b)
\end{center}
\caption{Diagrams for fluctuation contribution to the effective action
responsible
for Kondo screening corrections to magnetic (a) and spin liquid (b)
correlations.}
\label{polar}
\end{figure}
Here the external wavy lines stand for 
"semi-bosonic" field $\phi$ describing Kondo correlations (see \cite{ft1}).
These "semi-bosonic" fields are still bosons from the point of view of 
the permutation relations, but unlike true bosonic fields they do not satisfy
symmetric boundary conditions, and cannot condense in a state with zero 
frequency and momentum. So the Popov-Fedotov formalism gives an  adequate
description of the fact that there is no broken symmetry corresponding to 
Kondo temperature \cite{tw}. The polarization loop is formed by conduction 
electron propagator $G_0$ (solid line) and local semi-fermionic Green's function 
$D_m$ given by Eq. (\ref{2.18}) (see Appendix I for the explicit form of
these Green's functions). As a result the modified effective action is
\begin{equation}
{\cal A}_{\phi}=2\sum_{{\bf q},n}
\left[\frac{1}{\widetilde J}-\delta \Pi({\cal N})\right] 
|\phi_n({\bf q})|^2
\label{3.1}
\end{equation}
The logarithmic renormalization of the coupling constant is already 
taken into account in $\widetilde J$. Therefore the dimensionless 
integral $\delta \Pi$ includes only contribution due to nonzero magnetic 
molecular field \cite{am}, 
\begin{equation}
\delta \Pi({\cal N})=\left[
\frac{\pi}{2}\left(\frac{1}{\cosh(\beta{\cal N})}-1\right)
+O\left(\frac{{\cal N}^2}{T\epsilon_F}\right)\right]~.
\label{3.2}
\end{equation}
(see Appendix II for detailed calculations). 
This correlation correction should be incorporated in the equation for the 
free energy, so that 
\begin{equation}
\beta{\cal F}({\cal N})= \beta{\cal F}({\cal N}, 0) + 
{\rm Tr}\ln \left[\frac{1}{\widetilde J}-\delta \Pi({\cal N})\right]~. 
\label{3.3}
\end{equation}
Then differentiation of Eq. (\ref{3.3}) over the Neel order parameter ${\cal N}$, 
give the following self-consistent equation in the vicinity of renormalized 
transition point,
\begin{equation}
{\cal N}=\tanh\left(\displaystyle\frac{\beta I_{\bf Q}{\cal N}}{2}\right)
\left[\displaystyle
1-\frac{a_N}{\displaystyle\ln\left(T/T_K\right)}
\frac{\cosh^2(\beta I_{\bf Q}{\cal N}/2)} 
{\cosh^2(\beta I_{\bf Q} {\cal N})}\right]
\label{3.4}
\end{equation}
instead of Eq. ({\ref{2.16a}). Here the Kondo temperature, $T_K$, is defined 
as a temperature where the coefficient in front of $|\phi_{n=0}|^2$ in Eq.
(\ref{3.1}), i.e. the 
function $\tilde{J}^{-1}-\delta \Pi({\cal N}),$ turns into zero.   
It is seen that the screening corrections near the Neel transition point are
negative, $\delta\Pi({\cal N}\to 0)=-a_N(\beta{\cal N})^2 < 0$, so that Kondo
screening effectively increases the magnetic free energy, and eventually 
the logarithmic local field corrections {\it reduce} the Neel temperature.
The results
of numerical solution of Eq. (\ref{3.4}) are shown by a bold line in Fig. 2. 
Inset (a) illustrates the reduction of $T_N$ in comparison with the bare mean 
field Neel temperature $T_{N}^0=z\varepsilon_F\alpha^2/2$, where $\alpha=J/\varepsilon_F$ is
dimensionless coupling constant for the Doniach's diagram. 

(ii) {\it Kondo enhancement of SL transition}. Now we assume ${\cal N}=0$
in Eq.(\ref{2.16}) and subsequent equations. Following the same lines as 
in preceding subsection, one obtains     
the modified effective action 
\begin{equation}
{\cal A}_{\phi}=2\sum_{{\bf q},n}
\left[\frac{1}{\widetilde J}-\delta \Pi(I_{\bf q}\Delta)\right] 
|\phi_n({\bf q})|^2
\label{3.5}
\end{equation}
instead of (\ref{3.1}), and the polarization integral with the use of the diagram 
(b) from Fig.1 is given by
\begin{equation}
\delta \Pi(I_{\bf q}\Delta)=\sum_{\bf k}
\left[\frac{1}{\cosh \beta(I_{\bf k}\Delta)}-1+ 
I_{\bf k}\Delta\tanh(\beta I_{\bf k}\Delta)\right]
\frac{1}{\xi^2_{\bf k+q}+ (\pi/2\beta)^2}~.
\label{3.6}
\end{equation}
instead of  (\ref{3.2}) (see Appendix II). Here $\xi_{\bf p}=\varepsilon_{\bf p}-\varepsilon_F$
is the dispersion law for conduction electrons near the Fermi surface. 
Inserting the corresponding corrections to the 
free energy, 
\begin{equation}
\beta{\cal F}(\Delta)= \beta{\cal F}(0, \Delta) + 
{\rm Tr}\ln \left[\frac{1}{\widetilde J}-\delta \Pi(I_{\bf q}\Delta)\right]~. 
\label{3.3a}
\end{equation}
one obtains the corrected self-consistent equation for $\Delta.$
When
deriving this equation, the spinon dispersion can be neglected 
since $\Delta \to 0$ in a critical point. Then one has
\begin{equation}
\Delta= -\sum_{\bf q} \nu({\bf q})\left[\tanh\left(\frac{I({\bf q})\Delta}{T}
\right)+a_{sl}\frac{I_{\bf q} \Delta}{T\ln(T/T_K)}\right]
\label{3.4a}
\end{equation}
Here $a_{sl}\sim 1$ is a numerical coefficient.  It is seen from (\ref{3.4a})
that unlike the case of local magnetic order, Kondo scattering favors transition
to a SL state, because this scattering means in fact involvement of
itinerant electron spin degrees of freedom into spinon dynamics. Mathematically,
enhancement arises because 
$\delta\Pi(I_{\bf q}\Delta\to 0)=a_{sl}(\beta I_{\bf q}\Delta)^2 > 0$, 
so that Kondo
"anti-screening" effectively decreases the magnetic free energy. 
The results
of numerical solution of Eqs. (\ref{3.3a}) and (\ref{3.4a}) are presented by circles in Fig. 2.
\begin{figure}
\begin{center}
  \epsfxsize110mm \epsfbox{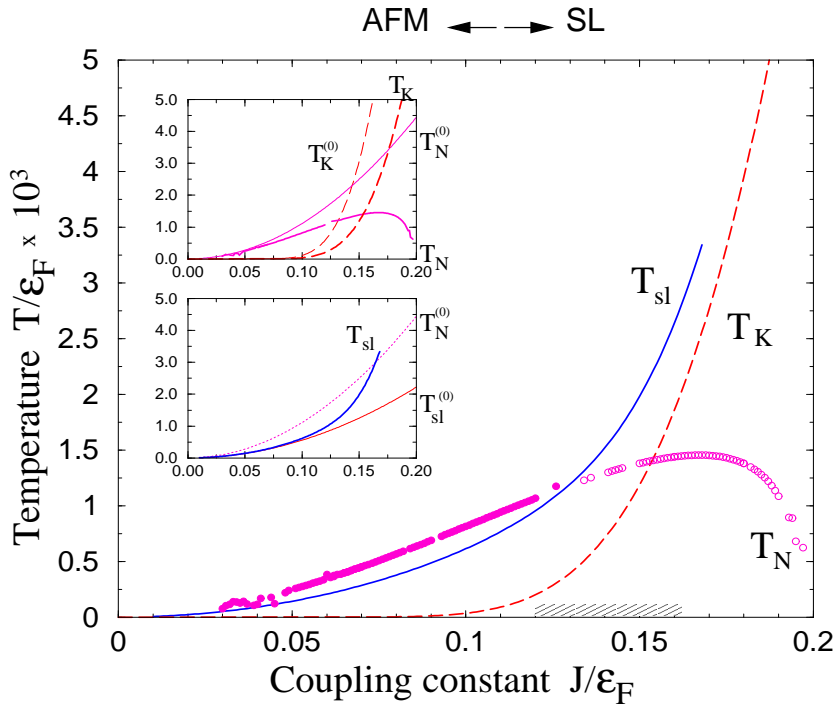}\hspace*{2cm}
\end{center}
\caption{Doniach diagram reconstructed due to Kondo screening
(see text for explanation).}
\label{don}
\end{figure}
Here filled circles correspond to the region where the AFM order overcomes the 
SL phase, and the light circles show unphysical "suppressed" AFM solutions 
obtained beyond the region of validity of  the mean-field equation (\ref{3.4}).
Two other characteristic temperatures, renormalized $T_K$ 
and $T_{sl}$, are shown by  dashed and solid line, respectively.  
The effects of suppression of $T_N$ (thin  and thick solid lines for bare and renormalized 
temperatures) and  $T_{K}$ (thin and thick dotted lines) are illustrated
by the upper and lower inset, respectively. 
As is seen from the modified Doniach's diagram, the interplay between three modes 
becomes significant in a critical region where the exchange   
coupling constant is close to the point  $\alpha_c=0.13$ where $I=\Delta_K$ in the
conventional Doniach's diagram. If the Kondo screening is not taken into account, then
$T_{sl}^{(0)}(\alpha)<T_{N}^{(0)}(\alpha)$ (thin solid and dotted lines in the lower inset).
The Kondo screening changes this picture radically, and as a result 
wide enough 
interval of the values of  parameter $\alpha$ just to the right of the critical
value $\alpha_c$ arises, where the enhanced 
transition temperature $T_{sl}$ exceeds both the reduced Neel temperature $T_N$
and the Kondo temperature $T_K$. 
The calculations of $T_{sl}$ presented in Fig. 2 are performed for $d=2$. The similar picture
exists for $d=3$, although the domain of stable SL state is more narrow (for a given 
value of $zI$).
This means that in this region the stable magnetic phase
is, in fact, the spin liquid phase. If one descends from high temperatures
in a hatched region of Doniach's diagram where $T_K \sim T_{N}$, 
the Kondo scattering suppresses the AFM correlations, but the SL correlations 
quench Kondo processes at some temperature $T_{sl}>T_K$. As a result the 
Kondo-type saddle point is not realized in the free energy functional in 
agreement with the  assumption used above in our derivation of Ginzburg-Landau expansion.  
The preliminary version
of this scenario was presented in  \cite{kkm94}. More refined 
mean field approach described here confirms and enhances this scenario,
however the SL liquid phase is still described in the mean field approximation. 
Although the local constraint for spin operators is not violated in 
Popov-Fedotov formalism, the gauge phase is still fixed \cite{lark}, so
the next task is consideration of fluctuation backflows described by the
higher order terms of Ginzburg-Landau expansion.  

\section{Ising spin glasses in Doniach's diagram}

In this section we consider the interplay between Kondo scattering 
and magnetic
correlations in case of {\it random} RKKY interaction (\ref{2.3}), where
the randomness results in formation of a spin glass phase. We consider
disorder induced by paramagnetic impurities in KL. As was shown in 
\cite{zus}, elastic scattering results in appearance of a random phase 
$\delta(r)$ in RKKY indirect exchange parameter,
\begin{equation}
I_{ij}\equiv I(r)\simeq - \left(\frac{J^2}{\varepsilon_F}\right)
\frac{\cos[2p_Fr -\frac{\pi}{2}(d+1)+\delta(r)]}
{(2p_F r)^d}
\label{4.1} 
\end{equation}
where $r=|R_i - R_j|$, $d$ is the dimensionality of  KL. This form of random  exchange 
predetermines two possible scenarios of SG ordering. 
\noindent\\
(i) Fluctuations take place
around a node of RKKY interaction (\ref{4.1}). 
This asymptotic is derived from the general equation for RKKY exchange parameter \cite{arist,stein}
\[
I_{ij}=-\frac{J^2}{\varepsilon_F} \frac{\pi}{d-1}\left(\frac{p_F a^2_0}{2\pi r}\right)^d
(p_Fr)^2\left[J_{d/2-1}(p_Fr)Y_{d/2-1}(p_Fr)+J_{d/2}(p_Fr) Y_{d/2}(p_Fr) 
\right] .
\]
($a_0$ is the lattice spacing, $J_\nu(x)$ and $Y_\nu(x)$ are the Bessel 
functions of 1st and 2nd kind). In this case FM and AFM 
bonds enter the partition function on equal footing, and quenched
independent random variables $I_{ij}$ can be described by Gaussian
distribution  $P(I_{ij}) \sim \exp (-I^2_{ij}N/(2I^2))$ \cite{sg}. 
The magnetic ordering effects also can be included in our approach
by introducing  nonzero standard deviation $\Delta I\neq 0$  into the
distribution $P(I_{ij})$ that, in turn, results in additional competition
between SG and AFM (or, in some cases, FM) states. Recently, the competition
between AFM and SG regimes was considered in \cite{osk}. 
\noindent\\
(ii) RKKY exchange fluctuates around some negative value in AFM domain
of exchange parameters. In this case there is a competition between SG,
SL and AFM phases. The third possibility, i.e. fluctuations in FM domain
is somewhat trivial because in this case Kondo fluctuations cannot 
significantly change freezing scenario.

We start with the case (i). To understand the situation qualitatively we make
following simplifying approximations. First, we consider only Ising-like
exchange in the Hamiltonian (\ref{2.3}):
\begin{equation}
H'=-\sum_{\langle ij \rangle}I_{ij}S^z_iS^z_j
\label{4.2}
\end{equation} 
This is usual approximation in the theory of spin glasses that allows one to forget
about quantum dynamics of spin variables \cite{ft4}. In the original paper  \cite{sege} the 
simplifying assumptions ($d=\infty$, separate electron bath for each localized spin) 
were made. Thus the form of spin-spin correlator was predetermined, and these 
assumptions allowed the authors to obtain exact solution in a framework of dynamical
mean field theory approach. We refrain from using these approximations. 
Second, we confine ourselves with the
mean field (replica symmetrical) solution of the Edward-Anderson (EA) model \cite{ea}.
This means that only a pairwise interaction of nearest neighbors is taken into 
account. The number $z$ of nearest neighbors should be big enough $(z^{-1}\ll 1)$ to verify
the mean field approximation. We consider the interplay between SG and Kondo-type
correlations by means of the replica method. We use the approach developed 
in  Ref. \onlinecite{kop} for the 
Sherrington-Kirkpatrick model \cite{shk}.
Both electron and semi-fermion
variables are replicated ($c \to c^a, f\to f^a$, where $a=1, \ldots , n$), and the number of replicas is tended to zero,
so that the free energy per cell is
given by the limit ${\cal F}=\beta^{-1}\lim_{n\to 0}(1-\langle {\cal Z}^n\rangle_{av})/(nN)$. Here the
replicated partition function is 
\begin{equation}
\langle {\cal Z}^n\rangle_{av}=\prod\int dI_{ij} P(I_{ij})\prod
D\{c^a_{i,\sigma} f^a_{i,\sigma}\}
\exp\left({\cal A}_0[c^a,f^a]-
\int_0^\beta d\tau H_{int}(\tau)\right)
\label{4.3}
\end{equation}
where ${\cal A}_0$ (\ref{2.9}) corresponds to noninteracting fermions.

Averaging over disorder and integrating out high-energy electronic states with the help of replica-dependent
Hubbard-Stratonovich trick one comes to the following equation 
\begin{equation}
\langle {\cal Z}^n\rangle_{av}=\prod\int D\{c^a,f^a,\phi^a\}
\exp\left({\cal A}_0+\frac{z I^2}{4N}{\tt Tr}[X^2] +
\int_0^\beta d\tau {\rm Tr}
\left\{\phi^a \bar c^a f^a +
\bar{\phi}^{a} \bar f^a  c^a  
-\frac{2}{\tilde J}|\phi^a|^2\right\}\right)
\label{4.4}
\end{equation}
with 
$$X^{ab}(\tau,\tau')=\sum_i\sum_{\sigma,\sigma'}
\bar f^a_{i,\sigma}(\tau)\sigma f^a_{i,\sigma}(\tau)
\bar f^b_{i,\sigma'}(\tau')\sigma' f^b_{i,\sigma'}(\tau').$$
Then following the standard pattern of replica theory for spin glasses \cite{rein,kop}
one fixes the saddle point in spin space related to EA order parameter $Q$. At this
stage the initial problem is mapped onto a set of independent Kondo scatterers for
low energy conduction electrons in external replica dependent effective magnetic field:
\begin{equation}
\langle Z^n\rangle_{av}=\exp\left(-\frac{1}{4}z (\beta I)^2N(n\tilde q^2+n(n-1) q^2)
+\sum_i\ln\left[\prod\int D\{f^a,\phi^a\}\int_x^G\int_{y^{a}}^G
\exp({\cal A}\{f^a,\phi^a,y^a,x\})\right]\right)
\label{4.5}
\end{equation}
where $\int_x^Gf(x)$ denotes
$\int_{-\infty}^{\infty}dx/\sqrt{2\pi} \exp(-x^2/2)f(x)$,
\begin{equation}
{\cal A}\{f^a,\psi^a,y^a,x\}=\sum_{a,\sigma}
\bar f^a_\sigma\left[(D_m^{(a)})^{-1}-\sigma h(y^a,x)\right]
f^a_\sigma-\frac{2}{\tilde J}\sum_n|\phi^a_n|^2
\label{4.6} 
\end{equation} 
and $h(y^a,x)=I\sqrt{zq}x+I\sqrt{z(\tilde q - q)}y^a$
is effective local magnetic field, which depends on diagonal, and 
off-diagonal elements of Parisi matrix, $\tilde{q}=\langle S_i^a(0)S_i^a(t \to \infty)\rangle$ 
and  $q=\langle S_i^a(0)S_i^b(t \to \infty)\rangle$ ($a\neq b)$, respectively. 
The latter one is the EA
order parameter $q_{EA}=q$.
Neglecting all fluctuations and retaining only first two terms in the exponent in Eq. (\ref{4.5}),
one comes to EA mean field equation for the free energy,
\begin{equation}
\beta{\cal F}=\frac{z(\beta I)^2}{4}\left[(1-\tilde{q})^2-(1-q)^2\right]-\int_x^G\ln(2 \cosh(\beta I x\sqrt{zq})).
\label{4.8}
\end{equation}
(see \cite{rein}). Then making the high temperature expansion, 
one obtains the Ginzburg-Landau equation in the vicinity
of SG transition \cite{bya}
\begin{equation}
\beta{\cal F}_{sg}=\frac{z(\beta I)^2}{4}q^2\tau_{sg}- c_{sg}q^3 +d_{sg}q^4
\label{4.9}
\end{equation}
where $\tau_{sg}=1-T_f/T$ and $T_f=\sqrt{z}I$ is a spin glass freezing temperature.
 
Like in the previous case of ordered KL we incorporate the
static replica dependent magnetic field $h$ in  semi-fermionic Green's functions. As a result, the 
modified effective action for Kondo fields arises like in Eqs. (\ref{3.1}) and (\ref{3.5})
\begin{equation}
{\cal A}[y^a,x]=\ln\left(2\cosh(\beta h(y^a,x)\right)
-\sum_n\left[\frac{1}{\tilde J}-\delta\Pi(h(y^a,x))\right]|\phi^a_n|^2
\label{4.7}
\end{equation}
Here similarly to Eq. (\ref{3.2})
\begin{equation}
\delta \Pi(h)=\left[
\frac{\pi}{2}\left(\frac{1}{\cosh(\beta h)}-1\right)
+O\left(\frac{h^2}{T\epsilon_F}\right)\right]~.
\label{4.10}
\end{equation}
Finally, performing Gaussian averaging over $\phi$ fields and taking the limit $n\to 0$ one
obtains the free energy 
\begin{equation}
\beta {\cal F}(\tilde q, q)=\frac{1}{4}z(\beta I)^2\left(\tilde q^2-
q^2\right)- \int_x^G\ln\left( \int_y^G2\cosh(\beta
h(y,x))/\left[1-J\Pi(0,h(y,x))\right]) \right). 
\label{4.11} 
\end{equation}
Corrected
equations for $q$ and $\tilde q$  are determined from conditions
$\partial {\cal F}(\tilde q, q)/\partial\tilde q=0$, $\partial {\cal F}(\tilde
q, q)/\partial q=0$. These are
\begin{eqnarray}
\frac{1}{2}z(\beta I)^2\tilde q & = &\int_x^G\frac{\partial
\ln{\cal C}}{\partial \tilde q},\;\;\;\; 
\frac{1}{2}z(\beta I)^2 q=-\int_x^G\frac{\partial\ln{\cal C}}{\partial q},\nonumber \\
{\cal C} & = & \int_y^G2\cosh(\beta h(y,x))/\left[1-J\Pi(0,h(y,x))\right])~.
\label{4.12}
\end{eqnarray}

Under the condition $ h(y,x) \leq 1$ a useful approximate equation
for ${\cal C}$ is obtained \cite{kop}:

\begin{equation}
\ln\left(C{\cal C}(x, \tilde q, q)\right)=\displaystyle
-\frac{1}{2}\ln\left(1+\gamma u^2(\tilde q - q)\right)+
\frac{u^2}{2} \frac{(\tilde q - q(1 + \gamma x^2))} {1+\gamma
u^2(\tilde q - q)} +\ln\left[ \cosh\left(\frac{u x
\sqrt{q}}{1+\gamma u^2(\tilde q - q)} \right)
\right]
\label{F} 
\end{equation}
Here the  following
short-hand notations are used: $u=\beta I\sqrt{z}$, $C=J/\epsilon_F\ln(T/T_K)$ and $\gamma=2c/\ln(T/T_K)$ with
$c=(\pi/4+2/\pi^2) \sim 1$. We note again that when $J=0$, which
corresponds to the absence of Kondo interaction, 
${\cal C}(x,\tilde q, q)= 2\exp\left(\frac{1}{2}z(\beta I)^2(\tilde q -
q)\right)\cosh(\beta I x\sqrt{zq})$, and standard EA equation
takes place, providing, e.g. an exact identity $\tilde q= 1$

In the vicinity of a freezing point Eq. (\ref{4.12}) acquires the form
\begin{eqnarray}
\tilde q & = & 1-\frac{2c}{\ln(T/T_K)} -
O\left(\frac{1}{\ln^2(T/T_K)}\right),
\label{4.13}\\
q & = & \int_x^G\tanh^2\left(\frac{\beta I x \sqrt{zq}} {1+2cz(\beta
I)^2(\tilde q - q)/\ln(T/T_K)}
\right)+O\left(\frac{q}{\ln^2(T/T_K)}\right) ~. 
\nonumber
\end{eqnarray}
As a result of numerical solution of Eqs. (\ref{4.13}) we obtain the analog of 
Doniach's diagram for a disordered KL where the spin glass freezing temperatures 
without and with Kondo screening contributions are shown ($T_f^{(0)}$ and $T_f$,
respectively). 
\begin{figure}
\begin{center}
  \epsfxsize110mm \epsfbox{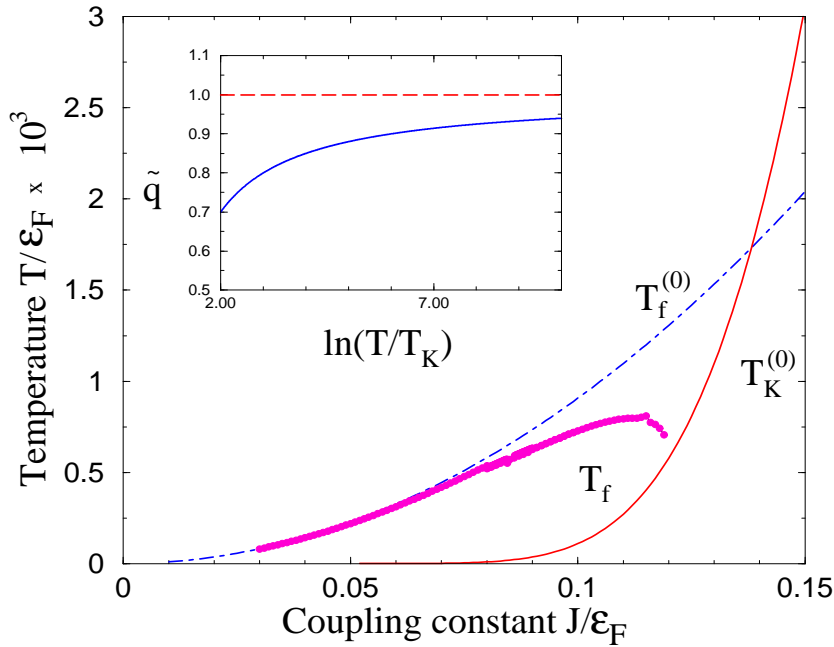}\hspace*{2cm}
\end{center}
\caption{Doniach's diagram for spin glass transition in disordered KL
(see text for explanation).}
\label{don2}
\end{figure}
Here $T_f^{(0)}$ is obtained from GL equation (\ref{4.9}) neglecting Kondo screening
effect, and $T_f$ was defined from Eqs. (\ref{4.13}) under additional condition 
$\partial ^2 {\cal F}_{sg}/\partial  q^2=0$. The influence of Kondo screening on  the 
diagonal element of Parisi matrix $\tilde{q}$ is illustrated by the inset (bare value of $\tilde{q}=1 $ 
is shown by the dashed line).  Like in the case of perfect KL, the screening
effect is noticeable when $T_f^{(0)} \sim T_K^{(0)}$. 

The influence of SG transition on a Kondo temperature for a KL with SG freezing was studied 
recently in Ref. \onlinecite{alba}. Although the Kondo effect in this paper is considered in 
a mean field approximation (i.e. Kondo screening  is treated as a true phase transition) and
a static ansatz was applied for SG, the authors obtained strong reduction of Kondo temperature
in the same region $T_K\sim T_f^{(0)}.$

\section{Correlations in Kondo lattice beyond the mean field}

The mean-field Doniach's diagram even in its improved form oversimplifies enormously the real
picture of interplay between three competing modes in effective action (\ref{2.11}).
First of all, the proximity of three characteristic temperatures, $ T_K$, $T_{sl}$ and $T_N$ means
that even when one of them is dominant, i.e. determines the local minimum of the free energy, 
two other define the fluctuations around the saddle point. Second, it is clear physically that
only Neel temperature, $T_N,$ is a temperature of a {\it real} phase transition, whereas
$T_{sl}$ and $T_K$ are merely  characteristic crossover temperatures. The main shortcoming of the
mean field approximation is that this approach treats all three modes on equal footing. The method
described in  preceding section allows one to get rid of artificial phase transition at $T=T_K$, however
the problems with description of SL phase still exist. 
Meanwhile,
it is known that the mean-field approximation for SL state
violates the local gauge invariance \cite{rvb,kkm97,lark,ba} and fixes the phase $\theta$ of SL mode 
$W$ (\ref{2.15a}). Second-order phase transition from paramagnetic to SL state \cite{ban} is an 
undesirable corollary of this crude approximation, and fluctuation corrections to the mean-field
solution cannot improve this defect of the theory. 

In this section we consider several scenarios of mode-mode correlations in a system 
described by the general
equation (\ref{2.8}),(\ref{2.11}) for effective action ${\cal A}$. First,
we offer the description of {\it crossover} to a SL state, which allows one to bypass the mean
field saddle point (\ref{2.16b}). It will be demonstrated that the interplay between fluctuations 
of the fields
$\phi$ and ${\bf Y_Q}$ can trigger the transformation of localized critical relaxation AFM modes into 
SL type correlations without loss of criticality. The main idea of our scenarios is that
the heavy fermion state of KL is, in fact, unconventional AFM state with spin excitations 
changing their character from Bose-like spin fluctuations or spin waves to Fermi-like spinon modes.
Next, we consider the
behavior of Kondo mode below $T_{sl}$ and describe the quenching of Kondo scattering by SL fluctuations
in a hatched part of Doniach's diagram (Fig. 2) where the static molecular field is absent.
\begin{figure}
\begin{center}
  \epsfxsize46mm \epsfbox{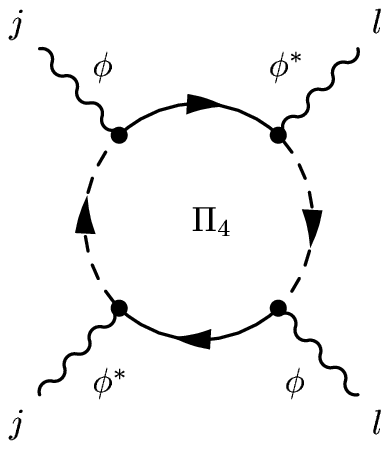}\hspace*{2cm}
  \epsfxsize46mm \epsfbox{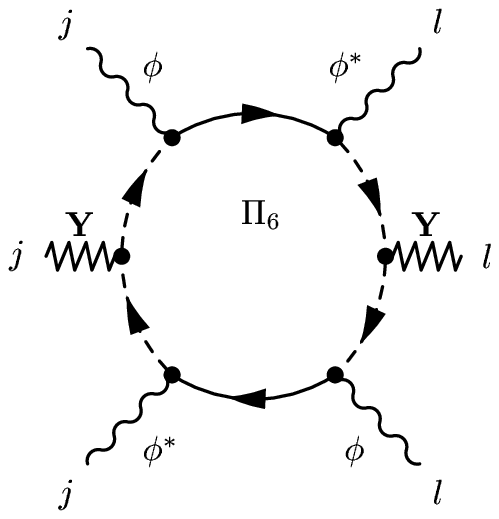}
a)\hspace*{7cm} b)
\end{center}
\caption{Diagrams for fourth and sixth order polarization operators $\Pi_4$ (a) and $\Pi_6$ (b) 
in effective action
responsible for mode-mode coupling.}
\label{polar2}
\end{figure}
We demonstrated above that the Kondo screening enhances SL correlations on a level of the mean field
approximation. Similar effect should exist on a more refined level of interacting fluctuation modes. 
To find the corresponding mechanism we refuse from bilocal representation of spin mode. Instead of
introducing the mode $W$ associated with gauge non-invariant $U(1)$ field described by the phase 
$\theta$ in Eq. (\ref{2.15a}), we consider the effect of interference of Kondo screening modes 
associated with spins located on different sites of KL. In fact we consider the high-temperature 
precursors of orthogonality catastrophe mentioned by Nozieres in his formulation of "exhaustion problem"
\cite{noz}. In a revised scheme we start with the action determined by the Hamiltonian (\ref{2.1}).
Starting with integration over "fast" electronic variables (with energies $\sim \varepsilon_F$), 
we obtain ${\cal A}_{int}$ in a form   
\begin{eqnarray}
{\cal A}_{int}  & = & -\frac{2}{\widetilde J}{\rm Tr}|\phi|^2 -
{\rm Tr}\frac{1}{I_{\bf q}}{\bf Y}_{\bf q}{\bf Y}_{\bf -q}    
-  {\rm Tr}\bar{f}_{j\sigma}\phi_j  G_0({\bf r}) \bar{\phi}_lf_{l\sigma}\nonumber \\
& - & {\rm Tr} \bar{\phi}_j\phi_l \Pi_4 \bar{\phi}_l \phi_j 
- {\rm Tr} {\bf Y}_j \bar{\phi}_j\phi_l \Pi_6 \bar{\phi}_l \phi_j {\bf Y}_l 
~.
\label{5.1}
\end{eqnarray}
Here instead of introducing the scalar mode $W$ we retained higher order terms in Kondo screening 
fields.
These terms are illustrated by the diagrams of Fig. 4  

The diagram of Fig. 4a describes interference of Kondo clouds around the sites 
${\bf R}_j$ and  ${\bf R}_l$. Zig-zag lines stand for AFM vector mode.
Like all screening diagrams in Fermi systems it contains Friedel-like oscillating factor.
To estimate the polarization operator we use the asymptotic form of the electron Green's function in d-dimension
at large distances \cite{kkm97a,arist}
\begin{equation}
G(r,\Omega)\sim \frac{1}{\displaystyle (p_Fr)^{\frac{d-1}{2}}}\exp \left[
-\frac{|\Omega|}{2\varepsilon_F} p_Fr + i\left(p_F r - \pi \frac{d+1}{4}
\right){\rm sgn} \Omega 
\right]
\label{5.1a}
\end{equation}
Inserting this function in a four-tail diagram of Fig. 4a, one comes to
\begin{equation}
\Pi_4 \sim - \frac{1}{T\varepsilon_F^2}\frac{\cos (2 p_F r - (d+1)\frac{\pi}{2})}{(2 p_F r)^{d-1}}
+ O\left(\frac{1}{\varepsilon_F^3} \ln\left[\frac{T}{\varepsilon_F}\right]
\right)
\label{5.2}
\end{equation}
Therefore we expect that this interference correlates with RKKY-type magnetic order,
and the interaction between the corresponding modes represented by the diagram (b) in Fig. 4 influences 
the magnetic response in a "critical" region of the Doniach's diagram. This response is determined by 
the fluctuation corrections to Neel effective action,
\begin{equation}
\delta{\cal A}_{eff}=\frac{1}{4}\sum_{{\bf q},\alpha,\omega_n}Y^\alpha({\bf q},\omega_n)
\left[I^{-1}({\bf q})+\chi_0\delta_{n,0}\right]
Y^\alpha({\bf q},\omega_n)
\label{5.3}
\end{equation}
Here $\alpha$ are Cartesian coordinates, $\chi_0=\beta/4$ is a static Curie susceptibility of isolated 
spin 1/2 (Fig. 5a). The term in square brackets is, in fact, inverse Ornstein-Zernicke correlator 
$\sim a_0^2({\bf Q-q})^2 + \tau_N$ at $T \gtrsim T_N$ and ${\bf Q-q} \to 0$. 
First non-vanishing correction 
to $\chi_0$ is given by Fig. 5b. In this diagram the spins $S_j$ and $S_j$ are screened independently,
(the wavy lines represent all parquet vertex insertions). In
the mean field approach the similar effects are described by Eq. (\ref{3.4}). 
\begin{figure}
\begin{center}
  \epsfxsize46mm \epsfbox{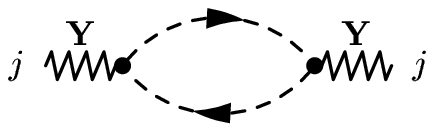}\hspace*{2cm}
  \epsfxsize46mm \epsfbox{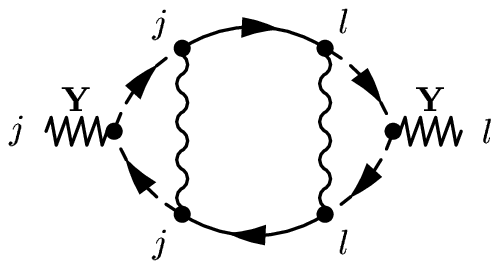}
a)\hspace*{7cm} b)
\end{center}
\caption{Diagrams describing local (Curie-type) magnetic susceptibility $\chi_0$ (a) and nonlocal 
correction taking into account Kondo screening of vertices (b).}
\label{polar3}
\end{figure}
\noindent
Indeed, each  vertex
correction $\Gamma_{i=j,l}(\omega, \epsilon)\sim \langle\phi(\epsilon)\bar{\phi}(\epsilon)\rangle$
gives the contribution $\sim 1/\ln (\epsilon/T_K)$, and integration over internal frequency $\epsilon$
results in $1/\ln (T_N^0/T_{K})$ correction in Eq. (\ref{3.4}) \cite{kkm94,kkm97a}.

The effects essentially beyond
the mean field are described by the diagrams that cannot be cut along a pair of electron propagators
(solid lines) (see  Fig. 6. and 7a).  
\begin{figure}
\begin{center}
  \epsfxsize46mm \epsfbox{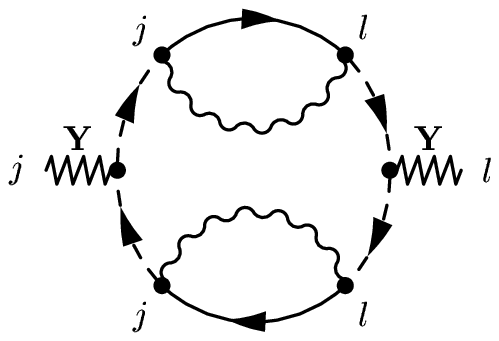}\hspace*{2cm}
  \epsfxsize46mm \epsfbox{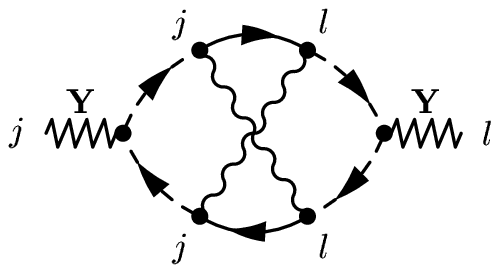}
a)\hspace*{7cm} b)
\end{center}
\caption{Leading diagrams describing interference of Kondo clouds in magnetic susceptibility (see
text for details).}
\label{polar4}
\end{figure}
\noindent
The first of these diagrams (Fig. 6a) can be treated as a nonlocal correction to one site spin susceptibility
(Fig. 5a) induced by interfering flow and counterflow of two Kondo clouds. 
As a result, the spin-fermion propagator becomes nonlocal without introducing the mean field order
parameter (\ref{2.15a}). 
The next diagram (Fig.6b) is a kind
of "exchange" by these clouds in the course of two-spinon propagation. Up to now we exploited the "proximity"
effects $T\gtrsim T_K$. A critical AFM mode given by the Fourier transform of the diagram of Fig. 5a 
with the wave vector ${\bf q\simeq Q}$ also exists in this temperature interval, and, moreover, this mode is 
dominant in spin susceptibility at $T\gtrsim T_N$. This means that the nonlocal contributions of 
Fig. 6 should be taken also at these ${\bf q}$. Due to nonlocality, the temperature dependence of spin
polarization loop will be slower than the Curie law  $1/T$, and the inverse static susceptibility given 
by these diagrams is   
\begin{equation}
\chi_{\bf Q}^{-1}(T)= \chi_0^{-1}(T)+ \chi_{sl}^{-1}(T)+\tilde I_{\bf Q}
\label{5.4}
\end{equation}
This deviation from Curie law results in delay of Neel phase transition or, in other words, in extension 
of critical regime to temperatures well below $T_N^0$ in accordance with scenario described in Ref. 
\onlinecite{kkm97}. Magnetic instabilities that can emerge at $T\ll T_N^0$ will be the instabilities of
spin liquid phase. These instabilities have much in common with itinerant fluctuational magnetism considered, 
e.g. in Refs. \onlinecite{mor,okun}.
\begin{figure}
\begin{center}
  \epsfxsize46mm \epsfbox{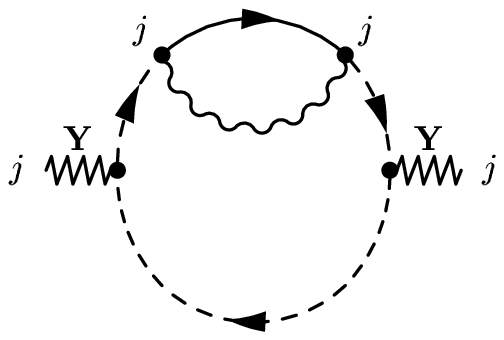}\hspace*{2cm}
  \epsfxsize46mm \epsfbox{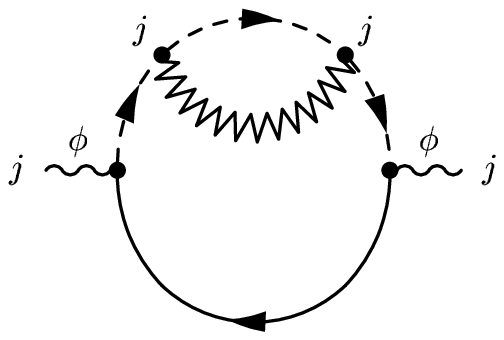}
a)\hspace*{7cm} b)
\end{center}
\caption{(a) Next to parquet approximation for Kondo correction to the magnetic 
susceptibility; 
(b) magnetic fluctuation correction to single site Kondo scattering.
}
\label{polar5}
\end{figure}

The diagram of Fig. 7a with bare spinon propagators gives only local correction to the susceptibility,
however at $T\ll T_N^0$ where the spinon lines are dressed by the self energies shown in Fig. 6a, 
this diagrams also becomes nonlocal and, therefore contributes in nonlocal term in the r.h.s. of Eq. (\ref{5.4}).
The processes taken into account in the diagram of Fig. 7b describe the feedback influence of spin 
fluctuations on the Kondo screening. This diagram together with higher order terms of the same type 
results in dynamical suppression of $T_K$ as a result of appearance of spin fluctuation energy 
$\omega_{sf}\sim \xi^{-z}$ in the Kondo logarithm. $\ln (\varepsilon_F/{\rm max}\{T, \omega_{sf}\})$.
This mechanism is effective not too close to real $T_N$ where the magnetic correlation length $\xi$ 
determining the short-range magnetic order is
still comparable with the lattice spacing (here $z$ is dynamical critical exponent). 

This schematic description is only scenario of the theory of critical phenomena in KL. 
We leave discussion of fluctuations around SG transitions beyond the scope of this paper.
Some details of  a new modulated replica symmetry breaking schemes, which combine tree and 
wave-like structures in AFM SG may be found in Ref. \onlinecite{osk}.
More detailed
calculation of critical magnetic and spin glass fluctuations in spin liquid will be published separately.

\section{Concluding remarks}
We derived in this paper the phase diagram for the Kondo lattice model, starting with a high temperature 
expansion of effective action. As a first step, we  succeeded in getting rid of one of fictitious saddle points, 
i.e. we avoided the introduction of "Kondo-condensate" averages 
$\langle c^\dagger_{k\sigma}f_{i\sigma}\rangle$
used in previous revisions of the Doniach's diagram \cite{rvb,coq}. In our modified Doniach's diagram
(Fig. 3) the {\it renormalized} $T_K$ is the lowest of all characteristic temperatures for all 
reasonable values of coupling constant $\alpha$ where one can neglect valence fluctuations. In fact,
the mean-field calculations of Ref. \onlinecite{coq} give  similar picture. The feedback
of this result is that the strong Kondo regime is unachievable in a critical region of Doniach's 
diagram, and the real role of Kondo screening for small 
$\alpha$ where $T_N>T_{sl}>T_K$ is to reduce localized magnetic moments and enhance the electronic
density of states around $\varepsilon_F$. Thus the moderately heavy fermion systems with relatively
big magnetic moments ordered antiferromagnetically arise $({\rm CeIn_3, CeAl_2})$ are possible examples)
\cite{ft2}. 

In a critical region of Doniach's diagram Kondo screening changes radically the behavior of
KL. According to our mean field results the conventional AFM order is suppressed at 
$T\sim T_{sl}\gtrsim T_N$.
The SL phase that arises instead is, nevertheless, close to magnetic instability, and one can expect that
spin subsystem eventually orders magnetically. If new transition temperature, 
${\widetilde T}_N$ is finite, 
the singlet spinon coupling is incomplete, so that RVBs have residual magnetic moments, and these moments 
are ordered at ${T=\widetilde T}_N$ (we emphasize once more  that $T_N$ marked by light circles  
in a hatched region of the phase diagram of Fig.3  is not a real transition temperature. It rather 
designates the temperature region where critical AFM fluctuations arise).
Of course, the magnitude of these moments is extremely small, and one can qualify 
this type of magnetic order is intermediate between localized and itinerant AFM. In the temperature 
interval
$\widetilde{T}_N<T<T_N$ the critical AFM relaxation mode characterizes 
the magnetic response of the system.   
When $\widetilde{T}_N=0$, one deals with quantum phase transition, and the case  $\widetilde{T}_N<0$,
apparently, corresponds to short-range correlations existing in a wide temperature interval $0<T<T_N$.
This picture describes in gross features the magnetic properties of magnetic KL, 
but any kind of quantitative
description will be possible only after realization of  scenarios for the critical behavior of 
spin liquid briefly sketched in Section V.  

Now we turn to discussion of conclusions that could be derived  from our theory concerning the nature of 
heavy fermion state. The most important one is that the separation of charge and spin degrees of
freedom existing in KL at high temperatures take place also in a strong coupling regime at $T\ll T_K$.
Indeed, at hight T exceeding all characteristic temperatures in KL the spin excitation spectrum
is simple structureless peak of the width $T$ around zero energy. This peak is manifested as Curie-type
magnetic susceptibility and trivial high-temperature corrections $\sim 1/T^n$ 
to all thermodynamic quantities due to weak paramagnetic spin scattering of conduction electrons,
whose Fermi liquid continuum exists as independent charge branch of elementary excitations. 
Since all transformations of spin subsystem occur at $T>T_K$ ( at least in a region of $\alpha <0.2$
where the valence fluctuations are still negligible), this central peak still exist in a strong
coupling regime. Below $T_{sl}\sim T_K$ this peak is formed by spin liquid excitations. The 
character of these excitations reminds relaxation modes in a picture of fluctuation 
itinerant magnetism \cite{mor,okun} in a wide temperature interval down to $T_{coh}$ where the
coherent spin liquid regime of Fermi type is established. The interaction between SL mode and 
conduction electrons is the same exchange-type scattering as at high temperatures. This coupling 
constant $\tilde{J}$ is, however, enhanced by the Kondo effect (see Eq. \ref{3.1}). The electrons in 
a layer of the width $T_K$ around Fermi level interact non-adiabatically with spin fermions at 
low $T$. As a result the giant Migdal effect arises \cite{kik} which results in strong electron 
mass enhancement. So, the heavy fermion state in accordance with this picture is a two-component
Fermi liquid where the characteristic energies of charge subsystem (slow electrons 
with $\epsilon< T_K$) and spin subsystem (spinons with $\omega \sim T_{sl}$) are nearly the same).

Exponentially narrow low-energy  peak of predominantly spin origin appears practically in all 
theories of strongly correlated electron systems. In the archetypal Hubbard model this peak 
arises on a dielectric side of Mott-Hubbard transition, and still exists on metallic side, where
the charge and spin degrees of freedom are already coupled. This is the point where the links
between Hubbard and Anderson models arise at least on a level of dynamical mean field theory 
(DMFT) valid at $d\to \infty$ \cite{kotl}. On the other hand the mean field solution that 
results in merged charge and spin degrees of freedom in a central peak becomes exact in 
the large-$N$ theories for $N=\infty$ saddle point \cite{col}. Recent achievements in this direction
are connected with confirmation of Noziere's prediction of 
second scale in Kondo lattice \cite{noz} 
in the limit of exhaustion regime of small electron concentration. At this temperature the "bachelor" 
spins form a coherent Fermi liquid and lose their localized nature. This anticipation was confirmed by
recent calculation within mean field slave boson approximation of $N\to \infty$ theory \cite{twos}.  
In our approach the regime of bachelor spins does not arise, because the 
Kondo coupling remains weak even at $T\ll T_K$ (see above), but the spin degrees of freedom become 
coherent at $T\sim T_{coh}$, so that two coherence scales is the intrinsic property of the model. 

Another aspect of large $N$ theories is the possibility of supersymmetric description that
allows combined description of spin degrees of freedom in a mixed fermion-boson 
$SU(N)$ representation 
\cite{pepin}. This approach allowed the authors to retain inter-site RKKY interaction in the limit
of $N\to \infty$ in spite of $1/N^2$ effect of suppression of all inter-site magnetic correlations
in a standard large $N$ approach. The use of Popov-Fedotov representation allows treatment of
different magnetic modes described by these operators as "semi-fermions" or "semi-bosons" in 
different physical situations \cite{ft1}. In this paper we appealed to $SU(2)$ symmetry.   
The general recipe of generation of modes with intermediate statistics between Fermi and Bose 
limiting cases for $SU(N)$ algebra is offered in \cite{kop01}.
In fact the eventual transformation of the states with intermediate statistics into true
fermions (bosons) occurs only at $T\to 0$. so this approach may be extremely useful for 
adequate description of quantum phase transitions \cite{ft3}.

In principle other collective modes can modify the scenario of AFM phase transition in KL. 
In particular, the low-lying crystal field excitations may intervene the magnetic phase transition
in the same fashion as Kondo clouds in our theory. Probably the CeNiSn family of semimetallic     
Kondo lattices is an example of such intervention \cite{cns}.

\section{Acknowledgments}
Authors thank A. Mishchenko for fruitful collaboration in the early stages of this work, 
A. Luther and G. Khaliullin  for valuable discussions and
A. Tsvelik for useful remarks. The work is supported by 
SFB-410. M. K. acknowledges the support of the Alexander von Humboldt Foundation, 
K.K. is grateful to Israeli-USA BSF-1999354
for partial support and to the University of W\"urzburg for hospitality.
\medskip\\
\begin{center}
\appendix{\bf APPENDIX I}
\medskip\\
\end{center}
\setcounter{equation}{0}
\renewcommand{\theequation}{AI.\arabic{equation}}

To evaluate the contribution of  Kondo mode  in the expansion (\ref{2.18b}) for effective
action, 
one needs the Fourier transform of the Green's function $K_{\phi}$ (\ref{2.18a}).
This is 
\begin{eqnarray}
\left(
\begin{array}{cccc}
\bar\phi_n({\bf k})G^0({\bf q})\phi_n({\bf k})& 0& \bar\phi_n({\bf k})G^0({\bf q})\phi_n({\bf k+Q})& 0\\
0&\bar\phi_n({\bf k})G^0({\bf q})\phi_n({\bf k}) & 0& \bar\phi_n({\bf k})G^0({\bf q})\phi_n({\bf k+Q}) \\
\bar\phi_n({\bf k+Q})G^0({\bf q})\phi_n({\bf k})&0&\bar\phi_n({\bf k+Q})G^0({\bf q})\phi_n({\bf k+Q})& 0\\
0& \bar\phi_n({\bf k+Q})G^0({\bf q})\phi_n({\bf k})& 0&\bar\phi_n({\bf k+Q})G^0({\bf q})\phi_n({\bf k+Q})
\end{array}
\right)
\label{a1.2}
\end{eqnarray}
The components $D_{m\sigma}({\bf q})$ of the semi-fermionic 
Green's function $D$ 
 in (20)
are determined by  inverting the matrix (18).
There are normal and anomalous components, 
\begin{equation}
-\int_0^\beta d\tau e^{i\omega_m \tau}
\langle T_\tau f_\sigma({\bf q},\tau)\bar f_\sigma({\bf q},0)\rangle=
\frac{i\omega_m-W_{\bf q}}{(i\omega_m-W_{\bf q})^2-Y^2}
\label{a1.1}
\end{equation}
and
\begin{equation}
-\int_0^\beta d\tau e^{i\omega_m \tau}
\langle T_\tau f_\sigma({\bf q},\tau)
\bar f_\sigma({\bf q}+{\bf Q},0)\rangle=
\frac{Y {\hat \tau}^z_{\sigma\sigma}}{(i\omega_m-W_{\bf q})^2-Y^2},
\label{a1.3}
\end{equation}
respectively. Here
$Y={\cal N}I_{\bf Q}/2$ and $W_{\bf q}=I_{\bf q}\Delta$.

To perform  calculations  in real space, one should know the
inverse Green's function (\ref{2.12}) in coordinate representation:
\begin{eqnarray}
D^{-1}_m(W,{\bf Y})=
\left(
\begin{array}{cccc}
i\omega_m+Y^z_j & Y^+_j        & W_{jl}         & 0\\
Y^-_j        & i\omega_m-Y^z_j& 0        & W_{jl}\\
W_{lj}       &   0       & i\omega_m-Y^z_l & Y^+_l\\
0       & W_{lj}        & Y^-_l           & i\omega_m+Y^z_l
\end{array}
\right)~.
\label{a1.5}
\end{eqnarray}
It should be noted 
that the nonlocal term  $W_{jl}$ in (\ref{a1.5}) responsible for SL correlations
transforms into diagonal term $W_{\bf q}$ in momentum representation 
(\ref{2.12}),
whereas
the local staggered field ${\bf Y}_i$ has non-diagonal matrix elements
in momentum space corresponding to AFM correlations at ${\bf q= Q}$.
\medskip\\
\begin{center}
\appendix{\bf APPENDIX II}
\medskip\\
\end{center}
\setcounter{equation}{0}
\renewcommand{\theequation}{AII.\arabic{equation}}
The sum of polarization integrals presented in Fig. 1 is given by the following equation
\begin{equation}
\Pi_n(Y, W_{\bf q})=-T\sum_{m,\sigma,{\bf p}}
D_{m\sigma}({\bf p})G^0_{m+n}({\bf p+q}) 
\label{a2.1} 
\end{equation}
Only normal component (\ref{a1.1}) survives in this equation as a result of spin summation.
The Neel loop (Fig.1a) after performing frequency summation acquires the form 
\begin{eqnarray}
\nonumber
\Pi(Y,0) & = & \sum_{{\bf p}}\left\{\tanh\left(\frac{\xi_{\bf p}}{2T}\right)
\left[\frac{\xi_{\bf p}-Y}{(\xi_p-Y)^2+\lambda^2}+
\frac{\xi_{\bf p}+Y}{(\xi_{\bf p}+Y)^2+\lambda^2}\right]\right.\\
\nonumber
& + & \frac{\lambda}{\cosh(Y/T)}\left[\frac{1}{(\xi_{\bf p}-Y)^2+\lambda^2}+
\frac{1}{(\xi_{\bf p}+Y)^2+\lambda^2}\right]\\
& - & \left.\tanh\left(\frac{Y}{T}\right)
\left[\frac{\xi_{\bf p}-Y}{(\xi_{\bf p}-Y)^2+\lambda^2}-
\frac{\xi_{\bf p}+Y}{(\xi_{\bf p}+Y)^2+\lambda^2}\right]\right\}
\label{a2.2}
\end{eqnarray}
Here $\xi_p=\varepsilon_p-\varepsilon_F,$ $\lambda=\pi T/2$. This integral is an even function of the order parameter, 
$\Pi(Y)=\Pi(-Y)$. Using the inequality
$Y\ll \epsilon_F$, two last terms can be simplified, and introducing the integral over the electron band with constant
density of states $\rho_0$, one has
\begin{eqnarray}
\nonumber
\Pi (Y,0)& = &\frac{1}{4}\rho_0\int_{-\epsilon_F}^{\epsilon_F} d\xi
\left\{\tanh\left(\frac{\xi_{\bf p}}{2T}\right)
\left[\frac{\xi_{\bf p}-Y}{(\xi_{\bf p}-Y)^2+\lambda^2}+
\frac{\xi_{\bf p}+Y}{(\xi_{\bf p}+Y)^2+\lambda^2}\right]\right\}\\
& + & \frac{\pi \rho_0}{2 \cosh(Y/T)} + \frac{\rho_0 Y}{\epsilon_F}
\tanh\left(\frac{Y}{T}\right)~.
\label{a2.3}
\end{eqnarray}
Incorporating $\rho_0$ in dimensionless variables, one has in the vicinity of Neel point where
$Y\ll T$
\begin{equation}
\Pi(Y,0)=\frac{1}{2}
\left(\ln\left(\frac{\epsilon_F^2}{4T^2}\right)+\pi\right)
-a_N\left(\frac{Y}{T}\right)^2+O\left(\frac{Y^2}{T\epsilon_F}\right)~.
\label{a2.4}
\end{equation}
The logarithmic term is, in fact, included in the renormalized coupling constant $\widetilde{J}$ in Eq. (\ref{3.1}) for 
effective action, and the remaining terms give Eq. (\ref{3.2}) for $\delta \Pi$.  
Deeper in magnetic phase where 
$Y\gg T$, the Kondo effect is quenched by molecular field, so that 
\begin{equation}
\Pi =
\ln\left(\frac{\epsilon_F}{Y}\right)
+b_N\left(\frac{T}{Y}\right)^2+O\left(\frac{T^2}{\epsilon_F^2}\right)
\label{a2.5}
\end{equation}
Numerical coefficients $a_N, b_N$ arising from approximate estimates of the integrals in Eq.(\ref{a2.3}) are 
of the order of unity. 

The SL loop (Fig. 1b) can be estimated for ${\bf q}=0$. After frequency summation it is presented by the following integral 
\begin{equation}
\Pi(0, \Delta)=\frac{1}{2}\sum_{\bf p}
\frac{\displaystyle \xi_{\bf p}\tanh\left(\frac{\xi_{\bf p}}{2T}\right)+
I_{\bf p}\Delta\tanh\left(\frac{I_{\bf p}\Delta}{T}\right)+
\frac{\lambda}{\displaystyle 2\cosh(I_{\bf p}\Delta/T)}}
{\xi_{\bf p}^2+\lambda^2}
\label{a2.6}
\end{equation}
This function is also even,
$\Pi(\Delta)=\Pi(-\Delta)$,  
Extracting from (\ref{a2.6}) the logarithmic term $\ln (\varepsilon_F/2T)$, one comes to Eq. ({\ref{3.6})
for $\delta \Pi$.
In a critical region of Doniach's diagram where
$\Delta \ll T$, one has   
\begin{equation}
\delta \Pi(0,\Delta) = 
a_{sl}\frac{\Delta^2}{2T^2}
\sum_{\bf p}\frac{\nu_{\bf p}^2}{\xi_{\bf p}^2+\lambda^2},
\;\;\;\;\;\;\;\;\;\;\;\;\;\;\;a_{sl}\sim 1. 
\label{a2.7}
\end{equation} 


\begin{thebibliography}{99}
\bibitem{ft1} 
   The specific feature of S=1/2 is that the set $2\pi T(m+1/4)$ can be
   interpreted both as "semi-fermionic" and "semi-bosonic" field. The 
   complementary field $\phi$ which appears in decoupling procedure for 
   bi-Grassmann products in representation (\ref{2.7a}) will be referred as 
   "semi-bosonic" field for the physical reasons, which are discussed below.
\bibitem{ft2} Metallic compounds containing U, apparently, should be described by the 
Anderson-lattice Hamiltonian 
because of partially itinerant nature of the 5$f$ electrons, so we refrain here from direct
application of our model to U-based heavy fermion systems.
\end{thebibliography}
\end{document}